\documentclass[lettersize,journal]{IEEEtran}

\usepackage{amsmath,amsfonts,bm}

\def\eqref#1{equation~\ref{#1}}
\def\1{\bm{1}}

\def\vh{{\bm{h}}}

\def\mA{{\bm{A}}}
\def\mB{{\bm{B}}}
\def\mC{{\bm{C}}}

\def\mI{{\bm{I}}}

\DeclareMathAlphabet{\mathsfit}{\encodingdefault}{\sfdefault}{m}{sl}
\SetMathAlphabet{\mathsfit}{bold}{\encodingdefault}{\sfdefault}{bx}{n}

\def\sR{{\mathbb{R}}}

\usepackage{amsmath,amsfonts}
\usepackage{amsthm}
\usepackage{algorithmic}
\usepackage{algorithm}

\usepackage{array}
\usepackage[caption=false,font=normalsize,labelfont=sf,textfont=sf]{subfig}
\usepackage{textcomp}
\usepackage{stfloats}
\usepackage{url}
\usepackage{verbatim}
\usepackage{graphicx}
\usepackage{cite}
\usepackage{academicons}
\usepackage{orcidlink}
\usepackage{hyperref}
\usepackage{subcaption}
\newtheorem{proposition}{Proposition}

\newtheorem{remark}{Remark}

\usepackage[normalem]{ulem}
\newcommand\tsout{\bgroup\markoverwith{\textcolor{red}{\rule[0.5ex]{2pt}{0.4pt}}}\ULon}

\begin{document}

\title{MMA: A Momentum Mamba Architecture for Human Activity Recognition with Inertial Sensors }

\author{\IEEEauthorblockN{
Thai-Khanh Nguyen \orcidlink{0009-0009-2028-9554}\textsuperscript{\textsection},
Uyen Vo \orcidlink{0009-0004-9785-7015}\textsuperscript{\textsection},
Tan M. Nguyen \orcidlink{0000-0003-0870-7054},
Thieu N. Vo \orcidlink{0000-0001-7957-5648},
Trung-Hieu Le \orcidlink{0000-0001-6323-6959},
Cuong Pham \orcidlink{0000-0003-0973-0889}}\\

\thanks{\textsuperscript{\textsection}These authors contributed equally.}
\thanks{Thai-Khanh Nguyen and Trung-Hieu Le are with the Falcuty of Information Technology, Dainam University, Hanoi University of Science and Technology, Hanoi 10000, Vietnam.}
\thanks{Cuong Pham and Uyen Vo are with the Faculty of Artificial Intelligence, Posts and Telecommunications Institute of Technology, Hanoi 10000, Vietnam.}
% \thanks{Trung-Hieu Le with the Falcuty of Information Technology, Dainam University, Hanoi University of Science and Technology, Hanoi 10000, Vietnam.}
% \thanks{Thieu N. Vo is with the Departments of Mathematics, National University of Singapore, Singapore 119076, Singapore.}
% \thanks{Cuong Pham is with the Faculty of Artificial intelligence, Posts and Telecommunications Institute of Technology, Hanoi 10000, Vietnam.}
\thanks{Thieu N. Vo is with the Department of Computer Science, University of Bath, Bath BA2 7AY, UK.}
\thanks{Tan Nguyen is with the Departments of Mathematics, National University of Singapore, Singapore 119076, Singapore.}
\thanks{Corresponding author: Cuong Pham (e-mail: cuongpv@ptit.edu.vn).}}

% \thanks{The code and dataset is available for research purpose and the link can be downloaded at https://github.com/bvvisual/CaRaCount.}
% \thanks{Digital Object Identifier 10.1109/TPAMI.2025.3548131}



\maketitle

\begin{abstract}
Human activity recognition (HAR) from inertial sensors is essential for ubiquitous computing, mobile health, and ambient intelligence. Conventional deep models such as Convolutional Neural Networks (CNNs), Recurrent Neural Networks (RNNs), and transformers have advanced HAR but remain limited by vanishing or exloding gradients, high computational cost, and difficulty in capturing long-range dependencies. Structured state-space models (SSMs) like Mamba address these challenges with linear complexity and effective temporal modeling, yet they are restricted to first-order dynamics without stable long-term memory mechanisms. We introduce Momentum Mamba, a momentum-augmented SSM that incorporates second-order dynamics to improve stability of information flow across time steps, robustness, and long-sequence modeling. Two extensions further expand its capacity: Complex Momentum Mamba for frequency-selective memory scaling. Experiments on multiple HAR benchmarks demonstrate consistent gains over vanilla Mamba and Transformer baselines in accuracy, robustness, and convergence speed. With only moderate increases in training cost, momentum-augmented SSMs offer a favorable accuracy-efficiency balance, establishing them as a scalable paradigm for HAR and a promising principal framework for broader sequence modeling applications.
\end{abstract}

\section{Introduction}
Human activity recognition (HAR) aims to identify human actions from sensor data automatically and supports applications in healthcare~\cite{Sigcha2023DeepLA, Nemati2021CoughBuddyMC, Gao2019TowardsRA, Bhattacharya2022EnsemHARAE}, human-computer interaction~\cite{Zhang2018FingerPingRF, Phinyomark2018EMGPR}, lifestyle monitoring~\cite{Jha2023AHA,Wang2019AttentionBasedCN}, sports analysis, and ambient assisted living~\cite{Sigcha2023DeepLA, Karim2024HumanAR}. Traditional HAR systems have relied on visual inputs such as RGB videos, depth maps, and skeletal data~\cite{Sun2020HumanAR, pham2021combining, Ghosh2022DeepLM}, but these approaches face challenges of occlusion, lighting sensitivity, computational cost, and privacy concerns~\cite{Sun2020HumanAR, Yang2024PrivacypreservingHA}.  

These limitations have driven a growing transition toward non-visual modalities, particularly inertial sensing through accelerometers and gyroscopes \cite{Lima2019Human}. Due to the mass adoption of smartphones and wearable devices, IMU data have become increasingly ubiquitous and accessible for HAR applications~\cite{Zhang2021DeepLI}. Inertial HAR offers unique advantages: privacy preservation, low power consumption, and fine-grained motion capture suitable for continuous and real-time monitoring ~\cite{Ignatov2018RealtimeHA, Qi2020ASA}. However, modeling inertial data introduces its own challenges. Sensor signals are inherently noisy, lack spatial context, and require robust temporal modeling of multivariate time series under resource-constrained conditions~\cite{Ahad2021, chen2021deeplearningsensorbasedhuman}.

Deep learning has become the dominant paradigm for inertial HAR. Convolutional Neural Networks (CNN) extract local temporal patterns~\cite{zeng2014convolutional}, whereas Recurrent Neural Networks (RNN) are designed to model sequential dependencies but often struggle to capture long-range temporal relationships due to issues such as vanishing gradients ~\cite{Farid2025ASA, Bengio1994LearningLD}. Hybrid CNN–RNN architectures have been developed to integrate the complementary strengths of both models~\cite{ordonez2016deep}. Transformer-based models~\cite{vaswani2017attention} have been introduced to capture global contextual dependencies through self-attention mechanisms, with HAR-specific extensions such as GAFormer~\cite{10317315} and MAMC~\cite{10705364}, but their quadratic computational and memory complexity causes the difficulty for real-time deployment.  

Structured state-space models (SSMs) offer a scalable alternative to Transformers, achieving linear-time complexity in sequence length. Among these, Mamba \cite{dao2024hungry} represents a significant advance, introducing an input-dependent selective mechanism that allows the model to dynamically focus on relevant information, with variants like HARMamba \cite{li2024harmamba} being adapted for HAR. Despite its effectiveness, Mamba's reliance on first-order dynamics can limit its ability to maintain stable gradient flow over long sequences, a critical challenge for noisy inertial data. The principle of incorporating second-order dynamics, inspired by momentum optimization methods \cite{polyak1964heavyball, Nesterov1983AMF, Sutskever2013OnTI}, offers a robust solution to this stability issue. This has been demonstrated in architectures such as MomentumRNN \cite{momentumRNN} and, more recently in the SSM context, by LinOSS \cite{rusch2025oscillatorystatespacemodels}, which uses oscillatory dynamics for stable long-range modeling. However, LinOSS employs time-invariant (i.e., static) state transition parameters that do not adapt to the input, thereby lacking the content-aware selectivity that is Mamba's key advantage. This exposes a clear research gap: no existing model integrates the adaptive selectivity of Mamba with the enhanced stability of second-order dynamics.  

{\bf{Contribution}}: We propose \textit{Momentum Mamba (MMA)} architecture, that augments Mamba’s dynamics with heavy-ball momentum. This integration enhances gradient stability, noise robustness, and temporal expressiveness while retaining linear scalability. Our contribution is three-fold:

\begin{enumerate}
    \item We introduce \emph{Momentum Mamba}, the first selective state-space model (SSM) that integrates momentum-driven second-order dynamics for improving gradient stability and temporal modeling in long-sequence inertial HAR.

    \item We theoretically prove that Momentum Mamba achieves better-structured spectrum than the Mamba baseline Mamba, explaining to the model’s stability enhancement.

     \item We demonstrate that the design principles of Momentum Mamba extend naturally to other advanced momentum-based optimization methods. In particular, we introduce Complex Momentum Mamba and Adam Momentum Mamba, which achieve frequency-selective memory via complex-valued momentum and adaptive control of momentum scaling, respectively.
\end{enumerate}

Extensive evaluations on a variety of inertial HAR benchmarks demonstrate that our models consistently outperform transformer, vanilla Mamba, and oscillatory SSM baselines, achieving superior trade-offs in accuracy, convergence speed, and resource efficiency.

\section{Preliminaries: Selective State Space Model}
 The classical state space model describes a continuous-time system that maps an input 
 The classical state space model describes a continuous-time system that maps an input 
$x(t) \in \mathbb{R}$ to an output $y(t) \in \mathbb{R}$ via a hidden state 
$\mathbf{h}(t) \in \mathbb{R}^{d \times 1}$, formulated as:
\begin{equation}
\label{eqn:cssm}
\begin{aligned}
    \vh'(t) &= \mA\vh(t) + \mB x(t), \\
    y(t) &= \mC\vh(t) + D x(t),
\end{aligned}
\end{equation}
where $\mA \in \sR^{d \times d}$, $\mB,\vh(t), \vh'(t) \in \sR^{d \times 1}$, $\mC \in \sR^{1 \times d}$, and $D \in \sR$. The \emph{structured state space sequence models (S4)} are inspired by  the classical SSMs and take the following discrete form of the continuous system in Eqn.~\ref{eqn:cssm}.
\begin{subequations}
    \begin{align}
        \vh_n &= \overline{\mA}\vh_{n-1} + \overline{\mB}x_n, \label{eqn:s4a} \\
        y_n &= \mC\vh_n, \label{eqn:s4b}
    \end{align}
\end{subequations}
where $\Delta \in \sR$ is a timescale parameter, and $\overline{\mA}, \overline{\mB}$ are discretized counterparts of $\mA, \mB$ via zero-order hold discretization. Specifically, $\overline{\mA} = \exp (\Delta \mA)$, and $\overline{\mB} = (\Delta \mA)^{-1}(\exp (\Delta\mA) - \mI)\cdot \Delta \mB \approx \Delta \mB$. Structured SSMs derive their name from the requirement that the matrix $A$, which governs the temporal dynamics, must adopt a specific structure to enable efficient sequence-to-sequence transformations suitable for deep neural networks. The initial designs introduced were the diagonal plus low-rank (DPLR) structure \cite{gu2022efficiently} and the purely diagonal structure \cite{gu2022parameterization,gupta2022diagonal,smith2023simplified}, with the latter remaining the most widely used.

% \emph{Selective State Space Models}, e.g., \emph{Mamba}~\cite{gu2024mamba}, improves S4 by selectively choosing to focus on or ignore inputs at every timestep. In particular, the parameters $\mB, \mC$, and $\Delta$ are set as the functions of $x_n$, thus becoming input-dependent parameters $\mB_n \mC_n$, and $\Delta_n$, respectively. As a results, the discretized parameters $\overline{\mA}_n = \exp (\Delta_n \mA)$ and $\overline{\mB}_n = \Delta_n \mB_i$ are also input-dependent. \emph{Mamba} chooses $\mB_n = \text{Linear}_{d}(x_n)$, $\mC_n = \text{Linear}_{d}(x_n)$, $\Delta_n = \text{softplus}(\theta + \text{Broadcast}_D(\text{Linear}_1(x_n)))$, where $\text{Linear}_d$ is a parameterized projection to a $d$-dimensional space, $\text{Broadcast}_D$ broadcasts a scalar into a $D$-dimensional vector, and $\theta \in \sR^{D}$ is a learnable parameter vector. 

\emph{Selective State Space Models} (SSMs), such as \emph{Mamba}~\cite{gu2024mamba}, extend S4 by selectively attending to or disregarding inputs at each timestep. Specifically, the parameters $\mB$, $\mC$, and $\Delta$ are defined as functions of the input $x_n$, yielding input-dependent forms $\mB_n$, $\mC_n$, and $\Delta_n$. Consequently, the discretized parameters also become input-dependent, with $\overline{\mA}_n = \exp(\Delta_n \mA)$ and $\overline{\mB}_n = \Delta_n \mB_n$. In \emph{Mamba}, these dependencies are instantiated as  
\begin{align}
\mB_n &= \text{Linear}_{d}(x_n), \nonumber \\
\mC_n &= \text{Linear}_{d}(x_n), \nonumber \\
\Delta_n &= \text{softplus}\!\left(\theta + \text{Broadcast}_D\!\big(\text{Linear}_1(x_n)\big)\right), \nonumber
\end{align} 
where $\text{Linear}_{d}$ denotes a learnable projection into a $d$-dimensional space, $\text{Broadcast}_D$ expands a scalar into a $D$-dimensional vector, and $\theta \in \mathbb{R}^D$ is a learnable parameter vector.

Compared to S4, \emph{Mamba} demonstrates better performance on information-dense data, such as language, particularly as the state dimension $d$ increases, thereby enhancing its information capacity ~\cite{li2024mamba, gu2024mamba}. However, Mamba is still restricted to first-order dynamics, thereby still having difficulty in capture long-range dependencies in long input sequences. Our Proposition~\ref{prop:mamba-limit} proved that Mamba indeed suffers from vanishing and exploding gradient issues. 
\begin{proposition}[Vanishing Gradients in Mamba]
\label{prop:mamba-limit}
Let $\{h_n\}$ be the hidden states of the Mamba architecture defined by
\[
h_n = \overline{A}_n h_{n-1} + \overline{B}_n x_n,
\]
where $\overline{A}_n = \exp(\Delta_n A)$ is diagonal with entries $e^{\Delta_n a_{n,i}}$ for $a_{n,i}<0$ and $\Delta_n>0$. Then the gradient of the loss $L$ with respect to $h_t$ satisfies
\[
\frac{\partial L}{\partial h_t}
= \frac{\partial L}{\partial h_T}\,\prod_{n=t+1}^{T} \overline{A}_n.
\]
If $\min_i a_{n,i}\ll 0$, then
\[
\Big\|\prod_{n=t+1}^{T}\overline{A}_n\Big\|\;\rightarrow 0 
\quad\text{as}\quad T-t\to\infty,
\]
so gradients vanish exponentially with sequence length.
\end{proposition}
\begin{proof}
By the chain rule,
\[
\frac{\partial L}{\partial h_t}
= \frac{\partial L}{\partial h_T}\cdot \frac{\partial h_T}{\partial h_t}.
\]
Unrolling the recurrence yields
\[
\frac{\partial L}{\partial h_t}
= \frac{\partial L}{\partial h_T}\cdot \prod_{n=t+1}^{T}\frac{\partial h_n}{\partial h_{n-1}}
= \frac{\partial L}{\partial h_T}\cdot \prod_{n=t+1}^{T}\overline{A}_n,
\]
since $\partial h_n/\partial h_{n-1} = \overline{A}_n$.  
Now, $\overline{A}_n$ is diagonal with entries $e^{\Delta_n a_{n,i}}$ where $a_{n,i}<0$. Thus,
\[
0 < e^{\Delta_n a_{n,i}} < 1 \quad \forall i,
\]
meaning each factor strictly contracts the corresponding gradient component. Repeated multiplication across $T-t$ steps yields exponential decay in the gradient magnitude. This aligns with the classical description of the \emph{vanishing gradient} \cite{bengio1994learning}, where Jacobians with eigenvalues strictly less than one in modulus drive gradients towards zero as the horizon grows.
\end{proof}
We further provide empirical evidence to validate Proposition~\ref{prop:mamba-limit} in Figure~\ref{fig:vanishing} below.

% Formally, the baseline Mamba recurrence is given by
% \begin{align}
%     h_n = \bar{\boldsymbol{A}}_n h_{n-1} + \bar{\boldsymbol{B}}_n x_n, \label{eq:mamba_update}
% \end{align}
% where $\bar{\boldsymbol{A}}_n = \exp(\Delta_n A)$ and $\bar{\boldsymbol{B}}_n = (\Delta_n A)^{-1}(\bar{\boldsymbol{A}}_n - I)\Delta_n B(x_n)$ are derived from continuous-time parameters. The step size $\Delta_n = \mathrm{softplus}(\mathrm{Linear}(x_n))$, input projection $B(x_n)$, and output projection $C(x_n)$ are input-dependent, while $A$ is a fixed base matrix.

\section{Related Work}
\label{sec:related_work}
This section reviews prior studies that form the foundation of our work. We organize the discussion into three parts: (A) the evolution of deep learning models for inertial HAR, (B) structured state-space models (SSMs) with a focus on Mamba and oscillatory extensions, and (C) momentum-inspired neural dynamics in sequence modeling. Together, these perspectives contextualize our proposed integration of momentum-driven recurrences into Mamba for robust inertial HAR.

\subsection{Deep Learning Models for Inertial HAR}

HAR using wearable inertial sensors such as accelerometers and gyroscopes is a central task in mobile health and ubiquitous computing. Traditional approaches relied on hand-crafted features and shallow classifiers such as decision trees, support vector machines, and naive Bayes~\cite{anguita2012human, liu2010research, deng2014cross}, but deep learning has enabled end-to-end representation learning from raw multivariate time series~\cite{wang2021deep}. 

CNN-based methods were among the earliest to show strong performance by capturing local temporal patterns directly from accelerometer and gyroscope signals~\cite{zeng2014convolutional,yang2015deep}. Lightweight variants further improved feasibility on embedded devices~\cite{abdulhay2021lightweight}. However, CNNs are constrained by limited receptive fields and often fail to capture long-term dependencies. To address this, RNNs such as LSTMs and GRUs were adopted to explicitly model sequential dynamics~\cite{hammerla2016deep}. Despite their effectiveness, these models suffer from vanishing gradients, slower training, and high memory consumption. Hybrid models such as DeepConvLSTM~\cite{ordonez2016deep} attempted to combine convolutional and recurrent layers, but the added complexity reduces suitability for real-time wearable systems.

Transformers introduced self-attention for global temporal modeling and have achieved state-of-the-art results in several HAR benchmarks~\cite{zhao2022tatt, li2023harformer}. To further adapt Transformers to inertial sensing, GAFormer~\cite{10317315} integrates Gramian Angular Field representations with graph alignment to capture structural dependencies in sensor sequences, while MAMC~\cite{10705364} introduces modality-aware attention for heterogeneous input fusion. Other variants such as LightFormer~\cite{huang2022lightformer} and HPFormer~\cite{lee2025hpformer} emphasize efficiency, with HPFormer in particular reducing self-attention complexity from $O(L^2)$ to $O(L \cdot \log L)$ to improve scalability in health informatics tasks. Despite these advances, the quadratic cost of vanilla self-attention and the high energy consumption of Transformer architectures remain major obstacles for real-time, resource-constrained HAR.

Beyond inertial-only methods, multimodal fusion combining inertial data with visual or depth inputs has been explored to improve robustness in noisy or occluded environments~\cite{koutrintzes2023multimodal, islam2023multimodal, yang2024crossmodal}. While these approaches enhance recognition performance, they are often limited by synchronization overhead, deployment cost, and energy consumption. By contrast, inertial sensors are inexpensive, ubiquitous, and energy-efficient, making them the most practical foundation for scalable, real-time HAR.

Overall, the trajectory of deep learning for inertial HAR spans CNNs, RNNs, and transformers, with recent work emphasizing lightweight and multimodal designs. However, a persistent challenge remains: balancing temporal modeling capacity with computational efficiency for resource-constrained deployment, motivating exploration of alternative paradigms such as structured state-space models.

\subsection{Structured State-Space Models and Mamba}

Structured state-space models (SSMs) provide an efficient alternative to attention by modeling sequence dynamics with linear-time recurrence. Early advances such as S4~\cite{gu2022efficiently} leveraged the HiPPO framework~\cite{gu2020hippo} to achieve long-horizon modeling with linear complexity, later refined by S5 and S6 for stability and precision. 

Mamba (S6)~\cite{dao2024hungry} further improved flexibility through input-dependent selective recurrence, combining linear scalability with content-aware dynamics. Variants such as HARMamba~\cite{li2024harmamba}, ActivityMamba~\cite{luo2025activitymamba}, and MHAR~\cite{le2025mamba} demonstrate their effectiveness for inertial and multimodal HAR, balancing accuracy with efficiency. 

Despite these advances, both Mamba and its variants remain constrained by first-order recurrences, which limit gradient regulation and robustness under noisy sensor streams. This motivates our exploration of momentum-augmented higher-order dynamics for inertial HAR.

\subsection{Momentum-Based Deep Learning Models}

Momentum, originally introduced in convex optimization~\cite{polyak1964heavyball}, accelerates convergence and stabilizes gradient trajectories, and its variants, such as Adan~\cite{xie2024adan} have been proven effective in large-scale deep learning. Beyond optimization, momentum has inspired neural architectures that embed second-order recurrences directly into model design. Examples include MomentumRNN~\cite{momentumRNN}, which augments hidden state updates with a velocity term, and NesterovNODE~\cite{nesterovNODE}, which reformulates neural ODEs as second-order systems for stability. Extensions such as physics-informed SSMs~\cite{chen2023physicsinformed} and generative SSMs for active inference~\cite{lanillos2020learning} further illustrate the versatility of momentum-like principles.  

LinOSS~\cite{rusch2025oscillatorystatespacemodels}exemplifies this trend by augmenting first-order SSMs with oscillatory second-order dynamics, achieving stability and efficiency for long-sequence modeling. However, existing designs employ either fixed oscillatory structures or remain outside the selective SSM framework. To address this gap, we propose \textit{Momentum Mamba}, which integrates heavy-ball momentum into Mamba’s input-dependent recurrence, enhancing gradient flow, noise robustness, and temporal expressiveness while preserving linear scalability.

\section{Momentum Mamba}
\label{sec:momentum_mamba}
\subsection{Incorporating Heavy-Ball Momentum into Mamba}
\begin{figure*}[htbp]
    \centering
    \includegraphics[width=1\textwidth]{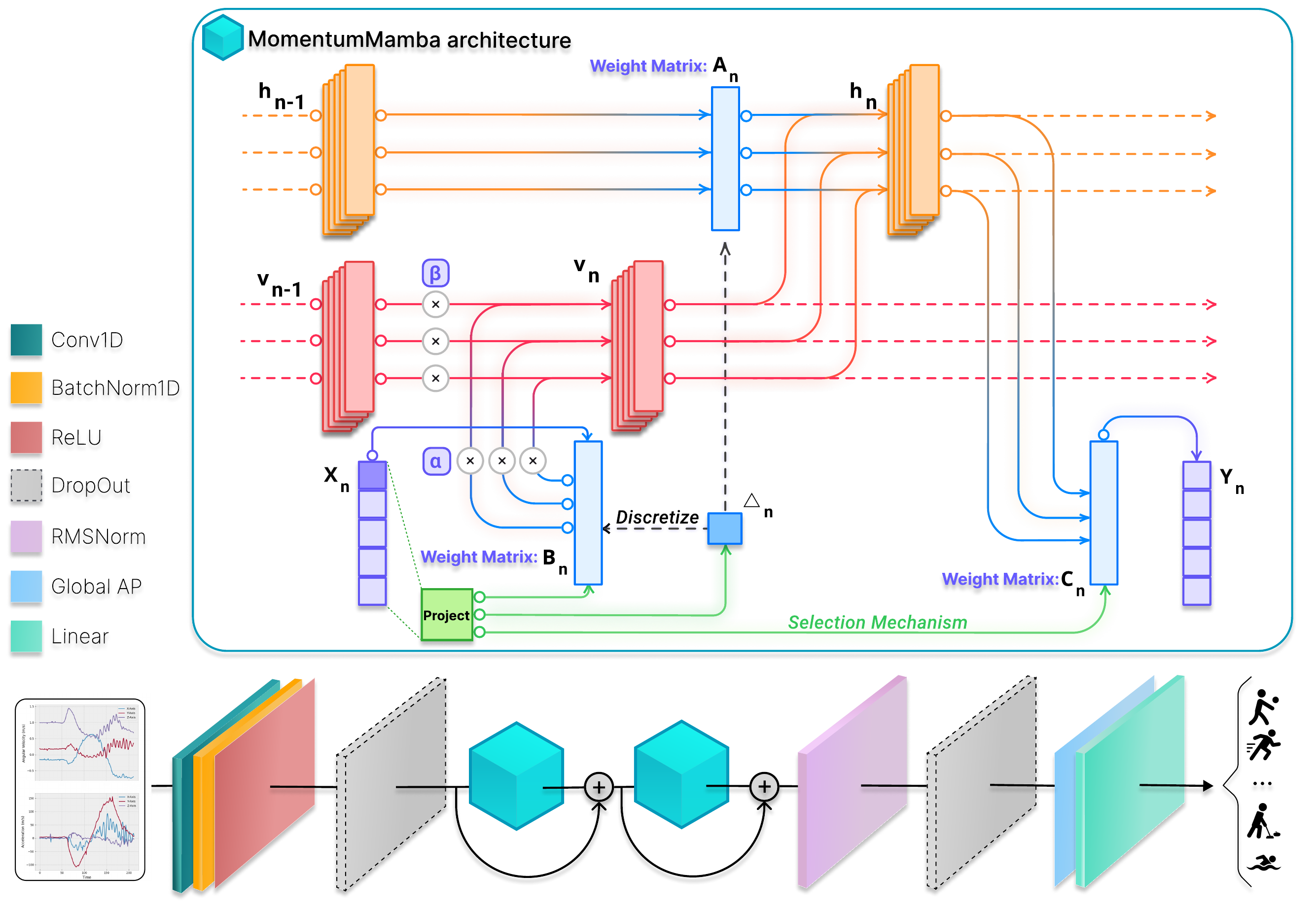}
    \caption{Overall architecture of the proposed \textit{Momentum Mamba} (MMA) framework for inertial HAR. 
    The pipeline consists of three main stages: a lightweight Conv1D front-end for local feature extraction, stacked Momentum Mamba layers for temporal modeling, and a compact classification head for activity recognition. 
    At its core, each Momentum Mamba block augments the standard Mamba recurrence with an auxiliary momentum state $v_n$ that accumulates input-driven updates through learnable parameters $(\alpha, \beta)$. 
    This dual-state design smooths high-frequency fluctuations, stabilizes long-range dynamics, and improves robustness while preserving the linear-time scan efficiency of structured state-space models.}
    \label{fig:momentum_mamba_arch}
\end{figure*}

At the core of the framework lies the \textit{Momentum Mamba} block, which extends the Mamba (S6) architecture--originally derived from continuous-time state-space models--with a momentum mechanism inspired by iterative optimization. In addition to the hidden state $h_n$, an auxiliary momentum state $v_n$ accumulates input-driven updates through learnable parameters $(\alpha, \beta)$. This dual pathway smooths high-frequency variations, stabilizes long-range dynamics, and improves gradient flow while preserving the scan efficiency of SSMs.  

Formally, the baseline Mamba recurrence is given by
\begin{align}
    h_n = \overline{\boldsymbol{A}}_n h_{n-1} + \overline{\boldsymbol{B}}_n x_n
\end{align}

where $\overline{\boldsymbol{A}}_n = \exp(\Delta_n A)$ and $\overline{\boldsymbol{B}}_n = (\Delta_n A)^{-1}(\overline{\boldsymbol{A}}_n - I)\Delta_n B(x_n)$ are derived from continuous-time parameters. The step size $\Delta_n = \mathrm{softplus}(\mathrm{Linear}(x_n))$, input projection $B(x_n)$, and output projection $C(x_n)$ are input-dependent, while $A$ is a fixed base matrix.

Drawing from the optimization analogy, we reinterpret $\overline{\boldsymbol{B}}_n x_n$ as a learned “gradient” and introduce a momentum accumulator to stabilize state evolution. 
We define an auxiliary momentum state $v_n \in \mathbb{R}^N$ and update the recurrence in three stages: 

\vspace{1mm} 
\noindent\textbf{(1) Momentum update:} 
\begin{align} v_n = \beta v_{n-1} + \alpha \overline{\boldsymbol{B}}_n x_n,
\end{align} where $\beta \in [0, 1]$ is a learnable momentum decay coefficient and $\alpha \in \mathbb{R}$ is a learnable step size. 

\vspace{1mm} 
\noindent\textbf{(2) Hidden state update:} 
\begin{align} h_n = \overline{\boldsymbol{A}}_n h_{n-1} + v_n, \end{align} so that the hidden state now evolves in response to a smoothed, temporally integrated signal rather than the raw instantaneous input. 

\vspace{1mm} 
\noindent
\textbf{(3) Output projection:} 
\begin{align} y_n = \boldsymbol{C}_n h_n,
\end{align} with $\boldsymbol{C}_n = C(x_n)$ derived from a learned projection conditioned on input.

% Formally, the Momentum Mamba block is defined as follows.

% \begin{definition}[Momentum Mamba]
% Let $\{x_n\}_{n=1}^L \subset \mathbb{R}^d$ be an input sequence. 
% A \emph{Momentum Mamba system} is a discrete-time dynamical system defined on the augmented state space 
% $\mathbb{R}^N \times \mathbb{R}^N$, with state variables $(h_n, v_n)$ evolving according to
% \begin{align}
%     v_n &= \beta v_{n-1} + \alpha \overline{\boldsymbol{B}}_n x_n, \label{eq:update_momentum}\\
%     h_n &= \overline{\boldsymbol{A}}_n h_{n-1} + v_n, \\
%     y_n &= \boldsymbol{C}_n h_n,
% \end{align}
% where $v_n \in \mathbb{R}^N$ is the \emph{momentum state}, $h_n \in \mathbb{R}^N$ is the \emph{hidden state}, 
% and $y_n \in \mathbb{R}^m$ is the output. 
% The parameters consist of a learnable momentum coefficient $\beta \in [0,1]$, a step size $\alpha \in \mathbb{R}$, 
% and input-dependent matrices 
% $\overline{\boldsymbol{A}}_n = \exp(\Delta_n A)$, 
% $\overline{\boldsymbol{B}}_n = (\Delta_n A)^{-1}(\overline{\boldsymbol{A}}_n - I)\Delta_n B(x_n)$, 
% $\boldsymbol{C}_n = C(x_n)$, 
% where $\Delta_n = \mathrm{softplus}(\mathrm{Linear}(x_n))$ and $A \in \mathbb{R}^{N \times N}$ is a fixed base matrix.
% \end{definition}

\begin{proposition}[Affine Recurrence Form]
\label{prop:parallel_temporal_smoothing}
The Momentum Mamba recurrence
\begin{align}
    v_n &= \beta v_{n-1} + \alpha \overline{\boldsymbol{B}}_n x_n, \label{eq:2} \tag{2}\\
    h_n &= \overline{\boldsymbol{A}}_n h_{n-1} + v_n, \label{eq:3} \tag{3}
\end{align}
admits an equivalent affine formulation
\begin{align}
    s_n = M'_n s_{n-1} + F'_n,
    \label{eq:momentum_mamba_affine}
\end{align}
where $s_n = \begin{bmatrix} h_n \\ v_n \end{bmatrix} \in \mathbb{R}^{2N}$ and
\begin{align}
    M'_n &= 
    \begin{bmatrix}
        \overline{\boldsymbol{A}}_n & \beta I \\
        0 & \beta I
    \end{bmatrix}, \qquad
    F'_n =
    \begin{bmatrix}
        \alpha \overline{\boldsymbol{B}}_n x_n \\
        \alpha \overline{\boldsymbol{B}}_n x_n
    \end{bmatrix}.
\end{align}
\end{proposition}

\begin{proof}
Stacking the hidden and momentum states into $s_n = [h_n^\top, v_n^\top]^\top$, 
the recurrence~\ref{eq:2}--\ref{eq:3} can first be written as the constrained linear system
\begin{align}
    T s_n = M_n s_{n-1} + F_n,
\end{align}
with
\[
T = \begin{bmatrix} I & -I \\ 0 & I \end{bmatrix}, \quad
M_n = \begin{bmatrix} \overline{\boldsymbol{A}}_n & 0 \\ 0 & \beta I \end{bmatrix}, \quad
F_n = \begin{bmatrix} 0 \\ \alpha \overline{\boldsymbol{B}}_n x_n \end{bmatrix}.
\]
Multiplying both sides by $T^{-1}$, which exists and is given by
\[
T^{-1} = \begin{bmatrix} I & I \\ 0 & I \end{bmatrix},
\]
yields the affine recurrence \eqref{eq:momentum_mamba_affine}.
\end{proof}

\begin{remark}[Parallelization and Temporal Smoothing]
Proposition~\ref{prop:parallel_temporal_smoothing} shows that the coupled hidden–momentum dynamics admit an affine recurrence. Intuitively, this means the entire update can be expressed as repeated applications of an affine map to the augmented state. 
Such maps compose associatively, as captured by
\[
    (a_1, a_2) \bullet (b_1, b_2) = (b_1 a_1,\, b_1 a_2 + b_2).
\]
Following the insights of LinOSS~\cite{rusch2025oscillatorystatespacemodels}, 
this associativity ensures that the sequence $\{s_n\}_{n=1}^L$ can be computed in $\mathcal{O}(\log L)$ parallel time with $\mathcal{O}(L \cdot N)$ total computation, making the formulation highly compatible with GPU-based parallelization.

Beyond efficiency, the inclusion of the momentum state $v_n$ acts as an exponential moving average, attenuating high-frequency input fluctuations. Thus, Momentum Mamba simultaneously preserves the scan efficiency of state-space models and enhances stability through temporal smoothing, without incurring additional asymptotic cost.
\end{remark}

Taken together, the formal results above establish Momentum Mamba as a natural extension of the Mamba framework: it retains the algebraic structure required for efficient parallelization while enriching the dynamics with a momentum pathway. This dual perspective--optimization-inspired smoothing combined with state-space discretization--provides the foundation for several practical advantages, which we summarize below.

\vspace{1mm}
\noindent\textbf{Selective dynamics.}
A defining feature of Mamba is its separation of dynamic modeling and input selectivity. The dynamics of the core transition is governed by the spectrum of the fixed matrix $A$, while the input dependence is injected through the learned functions $\Delta_n(x_n)$, $B_n(x_n)$, and $C_n(x_n)$. Our momentum-enhanced variant preserves this clean abstraction while inserting a smoothing buffer between the input and state transitions. The result is a more stable yet still content-aware recurrence.

\vspace{1mm}
\noindent\textbf{Computational efficiency.}
The introduction of momentum adds moderate overhead in terms of memory footprint, primarily due to the storage of augmented momentum states; however, the complexity remains linear in sequence length and is compatible with parallelization. The update rule remains linear in the hidden dimension and compatible with fast parallel scan operations. The parameter increase is negligible, involving only scalar momentum hyperparameters $\alpha$ and $\beta$.

\vspace{1mm}
\noindent\textbf{Expected benefits.}
By integrating momentum into the Mamba architecture, we aim to combine the selectivity and long-range expressivity of state-space models with the smooth convergence and robustness of optimization-based dynamics. We hypothesize that this formulation leads to the following:
\begin{itemize}
    \item More stable training across noisy or bursty input regimes;
    \item Improved convergence in long-horizon modeling tasks;
    \item Better generalization via reduced sensitivity to spurious input perturbations.
\end{itemize}
We refer to this enhanced architecture as \textit{Momentum Mamba}.

\subsection{Mitigating Vanishing Gradients in Momentum Mamba}
To address the inherent challenge of gradient vanishing in Mamba, we enrich the dynamics with an auxiliary momentum state $v_n$, yielding the augmented state $s_n = [h_n^\top, v_n^\top]^\top$. The momentum update
\[
v_n = \beta v_{n-1} + \alpha \overline{B}_n x_n
\]
acts as an exponential moving average of input signals, smoothing high-frequency variations. Embedding this update into the augmented recurrence produces the affine formulation of Equation~\eqref{eq:momentum_mamba_affine}, which fundamentally alters the Jacobian structure in backpropagation.

\begin{proposition}[Gradient Propagation in Momentum Mamba]
Let $s_n = [h_n^\top, v_n^\top]^\top$ evolve according to the affine recurrence
\[
s_n = M'_n s_{n-1} + F'_n,
\qquad
M'_n=\begin{bmatrix}\overline{A}_n & \beta I \\ 0 & \beta I\end{bmatrix}.
\]
Then the gradient of the loss $L$ with respect to $s_t$ is
\[
\frac{\partial L}{\partial s_t}
= \frac{\partial L}{\partial s_T}\,
\prod_{n=t+1}^{T} M'_n,
\]
which expands into the block form
\[
\frac{\partial L}{\partial s_t}
= \frac{\partial L}{\partial s_T}\,
\begin{bmatrix}
\prod_{n=t+1}^{T}\overline{A}_n &
\displaystyle\sum_{k = t+1}^{T}\!\Big(\prod_{n=t}^{T-k}\overline{A}_n\Big)(\beta I)^k\\[1ex]
0 & (\beta I)^{T-t+1}
\end{bmatrix}.
\]
\end{proposition}

\begin{proof}
The gradient propagation follows the chain rule:
\[
\frac{\partial L}{\partial s_t}
= \frac{\partial L}{\partial s_T}\cdot\frac{\partial s_T}{\partial s_t}
= \frac{\partial L}{\partial s_T}\cdot\prod_{n=t+1}^{T}\frac{\partial s_n}{\partial s_{n-1}}.
\]
Since $\partial s_n/\partial s_{n-1}=M'_n$, we obtain
\[
\frac{\partial L}{\partial s_t}
= \frac{\partial L}{\partial s_T}\prod_{n=t+1}^{T} M'_n.
\]
Expanding the product of block matrices yields the closed form. The lower-right block corresponds to repeated multiplication by $(\beta I)$, while the upper-right block captures convolution-like interactions between $\overline{A}_n$ and $(\beta I)^k$.
\end{proof}

\begin{remark}[Gradient Preservation by Momentum]
If $\beta\approx 1$, then $(\beta I)^{T-t+1}\approx I$ even for large horizons. Although the term $\prod_{n=t+1}^T \overline{A}_n$ may decay exponentially, the momentum pathway introduces eigenvalues close to unity. This spectral shift prevents the Jacobian product from collapsing to zero, ensuring that a non-negligible component of the gradient is preserved. As a result, Momentum Mamba maintains the gradient flow through the auxiliary state $v_t$, mitigating the vanishing gradient problem and improving the capacity to learn long-term dependencies.
\end{remark}

This result highlights that momentum in state-space models plays the same stabilizing role as in optimization: it accumulates and smooths gradient contributions while preventing their premature decay. Similarly to MomentumRNN \cite{nguyen2020momentumrnn}, the velocity state $v_n$ serves as a memory buffer that allows early inputs to exert a lasting influence on the final prediction.

\begin{remark}[Exploding Gradients]
While momentum alleviates vanishing gradients, overly large $\beta$ may lead to explosive gradients. In practice, gradient clipping \cite{pascanu2013difficulty} or normalization techniques can be applied to maintain numerical stability during training.
\end{remark}

\subsection{Beyond Heavy-Ball Momentum in Mamba: Complex and Adam Extensions}
\subsubsection{Complex Momentum Dynamics in Mamba}
To enrich the expressive power of state evolution, we extend Mamba with a second-order recurrence governed by complex-valued momentum:
\begin{align}
    v_n &= \beta v_{n-1} + \alpha \overline{\boldsymbol{B}}_n x_n, \label{eq:complex_momentum_state} \\
    h_n &= \overline{\boldsymbol{A}}_n h_{n-1} + v_n, \label{eq:complex_momentum_hidden} \\
    y_n &= \Re(\boldsymbol{C}_n h_n), \label{eq:complex_momentum_output}
\end{align}
where $\beta=\rho e^{i\theta}\in\mathbb{C}$ is the complex momentum coefficient whose magnitude $\rho$ controls damping and phase $\theta$ induces oscillations, $\alpha>0$ is the step size, $\overline{\boldsymbol{A}}_n=\exp(\Delta_n A)$ the input-dependent transition, and $\overline{\boldsymbol{B}}_n,\boldsymbol{C}_n$ are learned projections. The real part $\Re(\cdot)$ ensures that $h_n,y_n\in\mathbb{R}^N$.

This recurrence introduces oscillatory memory traces: each past input contributes with exponential decay $|\beta|^{n-k}$ and phase rotation $e^{i(n-k)\arg(\beta)}$. Unlike classical exponential smoothing, which only damps signals, complex momentum allows constructive or destructive interference in the complex plane, enabling the model to emphasize phase-aligned components.

\begin{remark}[Frequency-Aware Filtering]
When $|\beta|\approx 1$, low damping preserves oscillations whose frequency matches $\arg(\beta)$, while out-of-phase components cancel. Thus, Complex Momentum Mamba effectively learns frequency-selective filters, useful for oscillatory or quasi-periodic signals such as inertial sensor data or wave-like physical dynamics.
\end{remark}

\subsubsection{Adam Momentum for Mamba}
Complex momentum equips Mamba with spectral sensitivity; in contrast, Adam-style momentum introduces variance-aware adaptivity. Classical momentum methods smooth updates uniformly, but ignore per-dimension variability. Adam \cite{kingma2014adam} improves stability by tracking both first- and second-order moments, yielding adaptive coordinate-wise learning rates.

Inspired by MomentumRNN \cite{nguyen2020momentumrnn}, we embed this idea into Mamba:
\begin{align}
v_n &= \beta v_{n-1} + \alpha\, \overline{\boldsymbol{B}}_n x_n, \label{eq:adam_momentum_v}\\
m_n &= \gamma m_{n-1} + (1-\gamma)\,(\overline{\boldsymbol{B}}_n x_n)^2, \label{eq:adam_momentum_m}\\
h_n &= \overline{\boldsymbol{A}}_n h_{n-1} + \tfrac{v_n}{\sqrt{m_n}+\epsilon}, \label{eq:adam_momentum_h}\\
y_n &= \boldsymbol{C}_n h_n, \label{eq:adam_momentum_output}
\end{align}
where $v_n$ acts as a first-order momentum state with decay $\beta$ and step size $\alpha$, while $m_n$ tracks per-coordinate variance with decay $\gamma$. The hidden state update $v_n / (\sqrt{m_n}+\epsilon)$ thus adapts the effective step size based on local variance, stabilizing learning in noisy directions and accelerating it in sparse ones.

\begin{proposition}[Adaptive Recurrence Stability] 
Suppose inputs satisfy $\|\overline{\boldsymbol{B}}_n x_n\|\le B$ for all $n$. Then the normalized update $\frac{v_n}{\sqrt{m_n}+\epsilon} $ is uniformly bounded by $\tfrac{\alpha B}{\epsilon}$, ensuring bounded-input bounded-output stability of the hidden state $h_n$. \label{pro:adaptive} \end{proposition} 

\begin{proof} 
By definition $m_n \ge 0$ component-wise. Hence $\sqrt{m_n}+\epsilon \ge \epsilon$, giving \[ \left\|\frac{v_n}{\sqrt{m_n}+\epsilon}\right\| \le \frac{\|v_n\|}{\epsilon}. \] Since $v_n$ is an exponentially weighted sum of $\overline{\boldsymbol{B}}_k x_k$ scaled by $\alpha$, we have $\|v_n\|\le \tfrac{\alpha B}{1-\beta}$, and thus the normalized term is bounded by $\alpha B / \epsilon$. \end{proof}

\begin{remark}[Variance-Aware Updates]
The normalization in Eq.~\eqref{eq:adam_momentum_h} automatically reduces step size in high-variance directions while allowing faster adaptation in low-variance ones. Compared to standard Mamba, Adam Momentum Mamba improves stability under noisy or sparse inputs without altering the recurrence structure.
\end{remark}

\vspace{1mm}
\noindent\textbf{Discussion.}
Together, Complex Momentum Mamba and Adam Momentum Mamba enrich the dynamics of state-space models in complementary ways: the former introduces phase- and frequency-selective memory through complex-valued recurrences, while the latter provides variance-aware adaptivity via per-coordinate normalization. By combining spectral sensitivity with statistical stability, these extensions enhance both the representational power and the optimization robustness of Mamba for long-sequence learning. In the context of HAR, this dual design enables the model to better capture oscillatory sensor patterns while maintaining stable and adaptive training under noisy, heterogeneous input streams.

\section{Momentum Mamba Architecture for Inertial Human Activity Recognition}
\label{sec:mma-har}
% \begin{figure}%[htbp]
%     \flushleft
%     \includegraphics[width=0.9\columnwidth]{Images/architecture.png}
%     \caption{Overall architecture of the proposed MMA system. Six-axis inertial signals are processed by convolutional and normalization layers, followed by stacked Momentum Mamba blocks that capture both short- and long-range dependencies. The aggregated representation is mapped to activity classes through a global pooling and linear classifier.}
%     \label{fig:HAR_arch}
% \end{figure}

The overall architecture of the proposed \textit{Momentum Mamba} framework for inertial HAR is illustrated in Fig.~\ref{fig:momentum_mamba_arch}. The framework is designed to efficiently process synchronized six-axis inertial measurement unit (IMU) signals and to predict per-window activity labels through three main components: (\textit{i}) a convolutional front-end, (\textit{ii}) a momentum-augmented Mamba backbone, and (\textit{iii}) a lightweight classification head.

\subsection{Convolutional Front-End}
The first stage of the MMA employs a lightweight one-dimensional convolutional encoder that transforms raw six-axis inertial measurements (accelerometer and gyroscope) into high-dimensional temporal feature maps. A 1D convolutional layer with kernel size $3$ captures local temporal dependencies within short motion windows, followed by batch normalization, ReLU activation, and dropout for regularization. This design enhances local feature extraction and mitigates sensor noise, effectively projecting raw low-dimensional signals into a richer feature space for subsequent state-space processing.

The front-end is intentionally lightweight (about $1.8$K parameters for $6\!\rightarrow\!256$ dimensions), ensuring computational efficiency suitable for edge deployment. Its core purposes are threefold: (\textit{i}) capture local temporal patterns characteristic of human motion dynamics such as acceleration-deceleration transitions, (\textit{ii}) expand the representational capacity of sensor streams before entering the Momentum Mamba backbone, and (\textit{iii}) provide initial noise filtering through learned convolutions. Consequently, this module reduces the effective sequence length while enriching feature diversity, serving as an efficient bridge between raw IMU inputs and the momentum-augmented state-space layers that model long-range temporal dependencies in human activity recognition.

\subsection{Momentum Mamba Backbone}
The proposed framework extends the standard Mamba (S6) backbone by integrating a momentum mechanism inspired by second-order optimization dynamics. This modification, detailed in Section~IV of the Momentum Mamba formulation, introduces an auxiliary velocity state alongside the hidden state, forming a dual-state recurrence that enhances long-range temporal modeling and stabilizes gradient propagation.

The momentum-augmented update incorporates a smoothed velocity term that accumulates temporal information over time. This design reshapes the Jacobian spectrum, mitigating exponential gradient decay and improving robustness against high-frequency sensor noise--all while preserving the linear-time complexity of SSMs. As a result, MMA achieves smoother hidden-state evolution, stronger resilience to transient perturbations, and more stable convergence across long sequences.

The momentum mechanism provides a physically grounded inductive bias well suited for inertial signals: 
\begin{itemize}
\item \textbf{Inertial consistency:} The auxiliary velocity state captures the gradual, inertia-driven nature of human motion. 
\item \textbf{Noise robustness:} Temporal smoothing through momentum decay filters sensor noise while preserving activity-level dynamics. 
\item \textbf{Smooth transitions:} The model naturally represents gradual activity changes (e.g., walking $\rightarrow$ standing). 
\item \textbf{Extended dependencies:} The accumulated momentum state enables effective modeling of long-duration activities. 
\end{itemize}

Despite the additional momentum pathway, MMA retains linear-time complexity in sequence length $L$. Parallel scan (prefix-sum) implementations allow efficient computation of both $\mathbf{v}_t$ and $\mathbf{h}_t$, preserving GPU-friendly execution. Compared to vanilla Mamba, the overhead in training time and VRAM usage is marginal ($<5\%$ increase).

\subsection{Classification Head}
The final stage of the architecture aggregates the temporal representations obtained from the Momentum Mamba backbone into compact embeddings for activity classification. As formulated in Section IV, the hidden representations $\mathbf{h}_t$ produced at each time step are averaged along the temporal dimension to form a global feature descriptor, which is subsequently projected through a fully connected linear layer to produce class logits for $C$ activity categories.

This simple yet effective design introduces minimal additional parameters (less than 0.5\% of the total), ensuring a favorable trade-off between accuracy and efficiency. The use of global average pooling provides a holistic summary of temporal dynamics, while the linear projection enables direct mapping from latent representations to activity labels.

\subsection{Summary}
In summary, the proposed \textit{Momentum Mamba} framework establishes a principled and efficient approach to inertial human activity recognition by combining lightweight convolutional encoding, momentum-augmented state-space modeling, and compact classification. The integration of a momentum mechanism within the Mamba backbone provides several key advantages for HAR: it captures both transient and sustained motion patterns, preserves temporal continuity inherent in human movements, and stabilizes gradient propagation over long sequences. These properties enable the model to robustly distinguish activities that share overlapping short-term dynamics (e.g., walking vs.\ jogging) and to recognize complex behaviors that unfold over extended durations (e.g., cooking, exercising).

Furthermore, by maintaining linear-time complexity and a compact parameter footprint, MMA is well suited for deployment on wearable and mobile devices where computation and memory are constrained. The convolutional front-end effectively filters sensor noise and emphasizes local temporal structure, while the dual-state momentum backbone enriches temporal context through smooth, inertia-aware state transitions. Together, these components yield a model that is both physiologically interpretable and computationally efficient, advancing the robustness, accuracy, and real-time applicability of IMU-based human activity recognition systems.

\section{Experiments}
\label{sec:experiments}
\subsection{Dataset}

\begin{figure*}[htbp]
\centering
\includegraphics[width=0.9\textwidth]{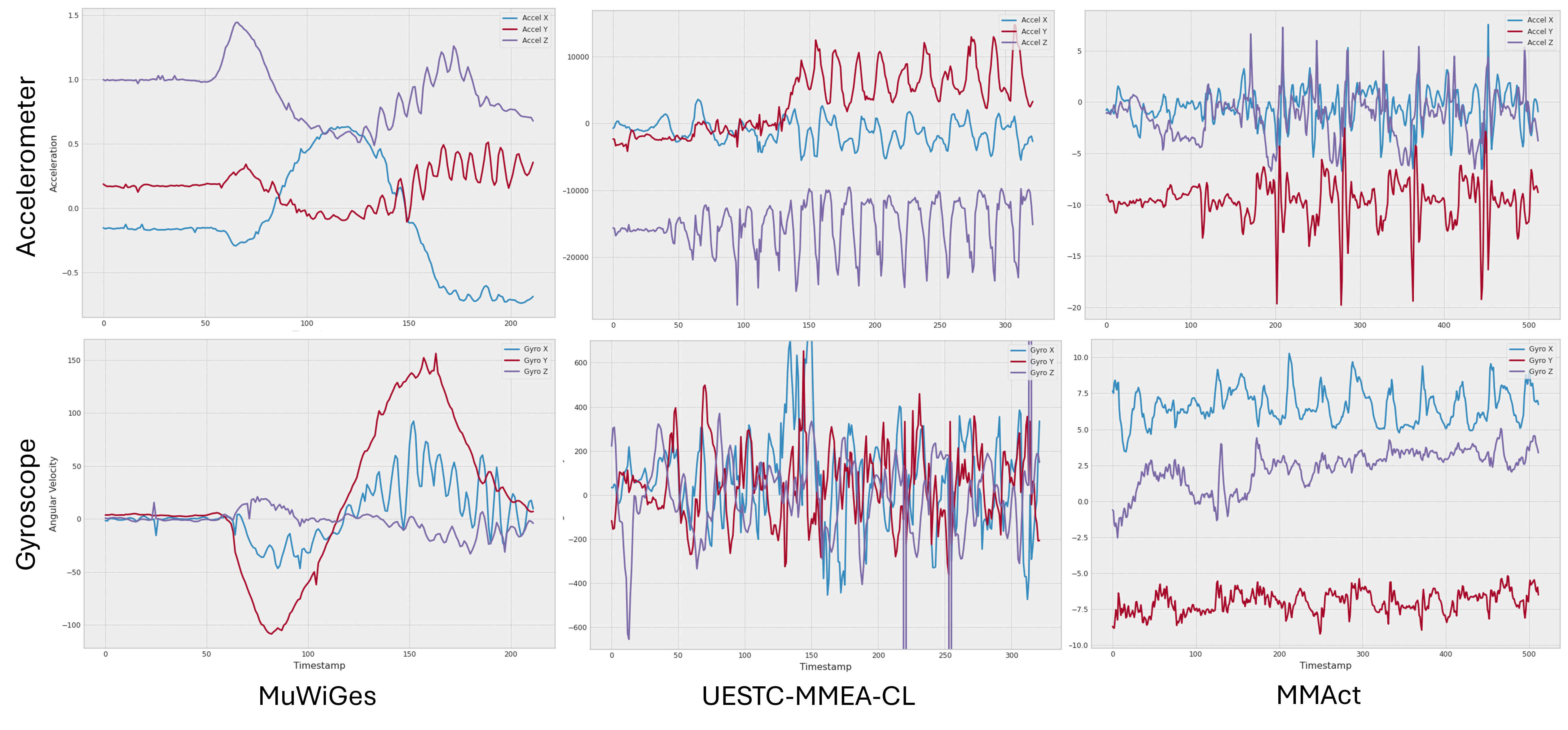}
\caption{Sensor signal examples from three benchmark datasets. The second and third rows display accelerometer and gyroscope signals, respectively, from the MuWiGes \cite{nguyen2023hand}, UESTC-MMEA-CL \cite{xu2023towards}, and MMAct \cite{kong2019mmact} datasets. These signals reflect the temporal variations of multi-axis motion data captured during human activities.}
\label{fig:dataset}
\end{figure*}

In this work, the proposed method is evaluated on three publicly available benchmark datasets: MuWiGes \cite{nguyen2023hand}, UESTC-MMEA-CL \cite{xu2023towards}, and MMAct \cite{kong2019mmact}. These datasets are designed for multimodal action recognition and provide synchronized RGB video along with inertial sensor data, including accelerometer and gyroscope signals. Fig. \ref{fig:dataset} illustrates sample data from each dataset, showcasing the integration of visual and motion modalities.

\begin{itemize} 

\item \textbf{MuWiGes dataset} is a carefully curated multimodal dataset acquired through a custom-designed wrist-worn device that integrates a high-resolution RGB camera and embedded IMU sensors. The camera records video streams at 1280×720 pixels with a frame rate of 30 FPS, while the inertial sensors comprising a tri-axial accelerometer and gyroscope sample motion data at 50 Hz. The dataset features gesture recordings from 50 participants (33 male and 17 female) spanning a broad age range (10–65 years), each performing 12 distinct hand gestures. Data collection was conducted across varied real-world environments, including homes and office spaces, to capture natural variations in background and user behavior. Each subject repeated each gesture spontaneously between 2 and 12 times, resulting in naturalistic gesture performance diversity. All gesture sequences were automatically segmented based on precise temporal boundaries indicating the onset and offset of each gesture. The final dataset comprises 5,408 synchronized samples that combine visual and inertial information. 

\item \textbf{UESTC-MMEA-CL dataset} dataset serves as a comprehensive multimodal benchmark tailored for evaluating continuous egocentric activity recognition. It comprises recordings of 32 distinct daily activities ranging from climbing stairs and drinking to shopping and playing cards performed by 10 participants across various environments, including both indoor and natural settings. Data were collected using smart glasses equipped with a first-person view camera and embedded IMU sensors. The camera records RGB video at a resolution of 640×480 pixels with a frame rate of 25 FPS, while the IMU captures motion data at 25 Hz. Each activity category includes approximately 200 synchronized samples, combining egocentric video and inertial measurements from accelerometer and gyroscope sensors.

\item \textbf{MMAct dataset}  is a multimodal dataset for activity recognition with 20 participants and 37 action classes. It includes seven synchronized modalities: RGB video, 2D keypoints, inertial (accelerometer, gyroscope), orientation, Wi-Fi, and barometric pressure. Data were collected using ceiling cameras (1920×1080, 30 fps), Google Glass (1280×720), and smartphones with inertial/context sensors (accelerometer 100 Hz, gyroscope 50 Hz), supplemented by smartwatches. Covering four environments, MMAct contains 35,084 labeled segments of diverse activities (e.g., typing, phone use, waving), making it a valuable benchmark for context-aware multimodal HAR.

\end{itemize}

\begin{table}[htbp]
    \centering
    \caption{Summary of three datasets used in experiments}
    \label{tab:datasets}
    \resizebox{\columnwidth}{!}{  
    \begin{tabular}{ccccc}
        \hline 
        \textbf{Datasets} & \textbf{MuWiGes~\cite{nguyen2023hand}} & \textbf{UESTC-MMEA-CL~\cite{xu2023towards}} & \textbf{MMAct~\cite{kong2019mmact}} \\
        \hline 
        \textbf{Activity type} & Hand gesture & Daily activity & Daily activity \\
        \textbf{Camera mounting} & Wrist & Head & Ambient/Head \\
        \textbf{IMU mounting} & Wrist & Head & Right wrist/thigh \\
        \textbf{Scenario} & Indoor (Home, Office) & Natural & Indoor (Office) \\
        \textbf{Data modalities} & RGB+Acc+Gyro & RGB+Acc+Gyro & RGB+Ori+Acc+Gyro \\
        \textbf{Total subjects} & 50 & 10 & 20 \\
        \textbf{Number of classes} & 12 & 32 & 37 \\
        \textbf{Total samples} & 5048 & 6522 & 35084 \\
        \textbf{Train/test splitting} & 3276/1772 & 4553/1316 & 28232/6930 \\
        \hline
    \end{tabular}
    }
\end{table}

Table \ref{tab:datasets} provides a summary of the three datasets tested in this study. These datasets differ in terms of the activities performed by the subjects, the placement of the cameras or IMU sensors, and the conditions under which the data was collected. In the MMAct dataset, cameras are positioned around the subjects to capture their actions from a front-facing view. Additionally, a camera mounted on Google Glass records egocentric videos, allowing a full view of the subject and their surroundings, which includes both hands and the area in front of them. In the second dataset, the camera is attached to the head, providing a view that captures both hands and the area in front, but lacks an additional observation perspective. In contrast, the final dataset, MuWiGes, has the camera mounted on the wrist, resulting in a more limited field of view. This restricted perspective makes gesture recognition from this angle significantly more challenging.

\subsection{Experimental Settings}
\textbf{Data preprocessing.} Raw inertial signals (tri-axial accelerometer and gyroscope) are resampled and segmented into fixed-length windows of $L=512$ with a 50\% overlap. Each channel is standardized to zero mean and unit variance using training statistics. Segments are then formatted as $X \in \mathbb{R}^{B \times L \times 6}$, where $B$ denotes the batch size.  

\textbf{Model configuration.} The MMA backbone consists of a lightweight Conv1D front-end, $N=2$ stacked Momentum Mamba layers with hidden size $d_{model}=128$, and a compact linear classification head. Dropout with probability 0.1 is applied throughout.  

\textbf{Training setup.} All models are trained end-to-end using categorical cross-entropy loss. Optimization is performed with Adam at an initial learning rate of $1\times10^{-3}$, weight decay of $1\times10^{-4}$, and cosine annealing learning rate scheduling. Mini-batches of size 16 are used for up to 50 epochs, with early stopping (patience = 10). Gradient clipping (global norm 1.0) is applied to stabilize training.  

\textbf{Hardware.} Experiments are implemented in PyTorch and executed on a single NVIDIA A30 GPU with 24 GB memory.  

\subsection{Experiment Results}
\subsubsection{Overview}
We evaluated Momentum Mamba and its variants on three public benchmarks: UESTC-MMEA-CL~\cite{xu2023towards}, MMAct~\cite{kong2019mmact}, and MuWiGes~\cite{nguyen2023hand}. Comparisons are made against both transformer-based models (Transformer, GAFormer~\cite{10317315}, MAMC~\cite{10705364}) and state-space approaches (Vanilla Mamba~\cite{dao2024hungry}, LinOSS~\cite{rusch2025oscillatorystatespacemodels}). Performance is measured in terms of accuracy, precision, recall, and F1-score, as summarized in Tables~\ref{table:muwiges_result}--\ref{table:mmact_result}.  

Across all datasets, Momentum Mamba consistently outperforms strong baselines, with average accuracy gains of $+2.77\%$ over Vanilla Mamba and $+1.64\%$-$19.11\%$ over Transformer. These improvements validate the effectiveness of embedding momentum-enhanced recurrence into the SSM backbone, yielding more stable gradient flow and stronger robustness to noisy inertial signals while preserving linear-time efficiency. 

Notably, Complex Momentum Mamba achieves the highest scores across all benchmarks, indicating that frequency-sensitive dynamics provide additional memory capacity beneficial for fine-grained motion patterns. This demonstrates that our momentum-driven framework is both effective in its base form and extensible to richer recurrence mechanisms.

\begin{table}[htbp]
    \centering
    \caption{Experimental results on the MuWiGes dataset. Bold values represent the best results.}
    \label{table:muwiges_result}
    \resizebox{\linewidth}{!}{%
    \begin{tabular}{lcccc}
        \hline
        Method & Accuracy (\(\uparrow\)) & Precision (\(\uparrow\)) & Recall (\(\uparrow\)) & F1-score (\(\uparrow\)) \\
        \hline
        Nguyen et al.~\cite{nguyen2023hand} & 95.60 & - & - & - \\
        GAFormer (2023)~\cite{10317315} & 98.33 & 98.25 & 98.33 & 98.30 \\
        Vanilla Transformer & 96.79 & 97.83 & 97.75 & 97.79 \\
        MAMC (2024)~\cite{10705364} & 96.20 & 96.56 & 95.90 & 96.21 \\
        Vanilla Mamba & 97.30 & 97.46 & 97.15 & 97.30 \\
        LinOSS (2025)~\cite{rusch2025oscillatorystatespacemodels} & 92.86 & 92.74 & 92.98 & 92.85 \\ \hline
        Momentum Mamba (Ours) & 98.43 & 98.56 & 98.26 & 98.41 \\
        \textbf{Complex Momentum Mamba (Ours)} & \textbf{98.64} & \textbf{98.67} & \textbf{98.52} & \textbf{98.59} \\
        \hline
    \end{tabular}
    }
\end{table}
\begin{table}[htbp]
    \centering
    \caption{Experimental results on the UESTC-MMEA-CL dataset. Bold values represent the best results}
    \label{table:uestc_result}
    \resizebox{\linewidth}{!}{%
    \begin{tabular}{lcccc}
        \hline
        Method & Accuracy (\(\uparrow\)) & Precision (\(\uparrow\)) & Recall (\(\uparrow\)) & F1-score (\(\uparrow\)) \\
        \hline
        Xu et al.~\cite{xu2023towards} & 59.70 & - & - & - \\
        GAFormer (2023)~\cite{10317315} & 79.00 & 78.41 & 79.94 & 79.15 \\
        MAMC (2024)~\cite{10705364} & 68.58 & 68.20 & 69.40 & 68.81 \\
        Vanilla Transformer & 73.21 & 73.50 & 72.90 & 73.17 \\
        LinOSS (2025)~\cite{rusch2025oscillatorystatespacemodels} & 77.73 & 76.20 & 78.12 & 77.14 \\
        Vanilla Mamba & 88.76 & 88.30 & 89.20 & 88.73 \\
        Mamba-LinOSS & 87.71 & 87.80 & 87.40 & 87.57 \\ \hline
        Momentum Mamba (Ours) & 92.32 & 92.10 & 92.60 & 92.39 \\
        \textbf{ComplexMomentumMamba (Ours)} & \textbf{94.07} & \textbf{94.20} & \textbf{93.95} & \textbf{94.07} \\
        \hline
    \end{tabular}
    }
\end{table}
\begin{table}[htbp]
    \centering
    \caption{Experimental results on the MMAct dataset. Bold values represent the best results.}
    \label{table:mmact_result}
    \resizebox{\linewidth}{!}{%
    \begin{tabular}{lcccc}
        \hline
        Method & Accuracy (\(\uparrow\)) & Precision (\(\uparrow\)) & Recall (\(\uparrow\)) & F1-score (\(\uparrow\)) \\
        \hline
        Multi-teacher (2019)~\cite{kong2019mmact} & 62.67 &--&--&--\\
        VLMs (2023)~\cite{tavassoli2023expanding} & 64.47 &--&--&--\\
        VSKD (2022)~\cite{ni2022cross} & 60.14 &--&--&--\\
        GAFormer (2023)~\cite{10317315} & 60.76 & 61.00 & 60.52 & 60.75 \\
        Vanilla Transformer & 59.91 & 59.46 & 60.40 & 59.91 \\
        Vanilla Mamba & 71.86 & 72.24 & 71.56 & 71.83 \\
        LinOSS (2025)~\cite{rusch2025oscillatorystatespacemodels} & 72.16 & 71.28 & 72.53 & 71.90 \\ \hline
        Momentum Mamba (Ours) & 75.49 & 75.86 & 75.10 & 75.45 \\
        \textbf{Complex Momentum Mamba (Ours)} & \textbf{76.62} & \textbf{76.80} & \textbf{76.45} & \textbf{76.63} \\
        \hline
    \end{tabular}
    }
\end{table}

\subsubsection{Dataset-specific performance}
\paragraph{MuWiGes} 
Momentum Mamba attains $98.43\%$ accuracy, exceeding Vanilla Mamba by $+1.13\%$ and slightly surpassing GAFormer. Gesture recognition from wrist-mounted inertial sensors is inherently difficult due to subtle motion cues, subject variability, and contamination from incidental hand jitter. Momentum Mamba addresses these challenges by smoothing transient fluctuations while retaining phase-consistent motion information, enabling reliable separation of gestures with highly similar dynamics (e.g., “rotate wrist” vs. “twist cap”). The Complex Momentum Mamba variant further improves accuracy to $98.64\%$, indicating that complex-valued recurrence enhances robustness against rapid oscillations and improves discrimination of fine-grained gesture classes. 

\paragraph{UESTC-MMEA-CL} 
On this fine-grained daily activity dataset, Momentum Mamba achieves $92.32\%$ accuracy, a $+3.56\%$ improvement over Vanilla Mamba. Recognition is difficult because many actions (e.g., “reading” or “typing”) produce weak, low-frequency signals that are easily obscured by noise or incidental fluctuations. By leveraging second-order recurrence, Momentum Mamba aggregates these subtle components across long horizons while attenuating irrelevant variations, enabling the model to maintain stable and discriminative temporal patterns. In addition, the Complex Momentum Mamba variant raises performance to $94.07\%$ accuracy and F1-score, suggesting that frequency-sensitive dynamics further sharpen the model’s ability to separate overlapping or weak activity signals--an advantage particularly evident for subtle or repetitive motion classes.

\paragraph{MMAct} 
The MMAct dataset, with 37 action classes under diverse conditions, remains one of the most challenging HAR benchmarks using inertial data. Signal variations arise not only from activity type but also from device placement, environment, and synchronization quality. Momentum Mamba reaches $75.49\%$ accuracy, improving upon Vanilla Mamba by $+3.63\%$ and Transformer by $+15.58\%$. Its advantage lies in stabilizing hidden state evolution, preventing overreaction to short-lived or conflicting signals across sensor axes, and thereby capturing persistent motion trends. For this dataset, the Complex Momentum Mamba variant pushes performance to $76.62\%$ accuracy and $76.63$ F1-score, suggesting that complex-valued dynamics are especially beneficial for recognizing composite or multi-step activities involving subtle temporal transitions.

Overall, the results across MuWiGes, UESTC-MMEA-CL, and MMAct confirm that Momentum Mamba provides consistent and substantial improvements over state-of-the-art baselines by stabilizing long-range temporal modeling of inertial signals. Moreover, the Complex Momentum Mamba variant further extends these benefits with frequency-sensitive dynamics, yielding additional gains in challenging scenarios that involve subtle, noisy, or multi-step activities.

\subsubsection{Efficiency analysis}
\begin{table}[h!]
\centering
\caption{Comparison of Mamba, MomentumMamba, and ComplexMomentumMamba $(d_{model}=128, n_{layers}=2, d_{state}=64, d_{conv}=4, expand=2).$}
\resizebox{\linewidth}{!}{
\begin{tabular}{lcccc}
\hline
\textbf{Model} & \textbf{Time (s)} & \textbf{FLOPs (MFLOPS)} & \textbf{Params (K)} & \textbf{VRAM (MB)} \\ \hline
Mamba & 0.0052 & 311.285 & 313.632 & 188.41 \\
MomentumMamba & 0.007 & 278.091 & 313.760 & 212.41 \\
CMMamba & 0.031 & 311.646 & 412.065 & 396.91 \\ \hline
\label{tab:efficient_annalysis}
\end{tabular}}
\end{table}

In addition to accuracy, practical deployment of HAR models depends heavily on computational efficiency and memory footprint. To this end, we compare Mamba, Momentum Mamba, and Complex Momentum Mamba under identical configurations $(d_{model}=128, n_{layers}=2, d_{state}=64, d_{conv}=4, expand=2)$, as summarized in Table~\ref{tab:efficient_annalysis}.

Vanilla Mamba offers the lowest latency, with an inference time of $0.0052$ seconds and a VRAM consumption of $188.41$ MB. Momentum Mamba introduces only a marginal overhead, increasing runtime to $0.007$ seconds and memory to $212.41$ MB. Interestingly, its FLOPs ($278.091$ MFLOPs) are slightly lower than that of Vanilla Mamba ($311.285$ MFLOPs), reflecting that the integration of momentum does not impose significant computational burden. This result confirms that the proposed recurrence can be incorporated with negligible cost while delivering substantial improvements in recognition accuracy.

On the other hand, Complex Momentum Mamba achieves the strongest performance but at a notable efficiency trade-off. Its runtime ($0.031$ seconds) and VRAM requirement ($396.91$ MB) are considerably larger, and the parameter count rises to $412$K compared to $314$K for both Mamba and Momentum Mamba. These statistics highlight that while the complex variant is highly effective for accuracy-sensitive applications, it demands significantly greater resources.

Overall, this comparison demonstrates that Momentum Mamba provides the best balance between efficiency and performance, making it suitable for real-world HAR scenarios where computational resources are limited. In contrast, Complex Momentum Mamba represents an extensible, high-accuracy alternative tailored to scenarios where precision is prioritized over efficiency.

\subsubsection{Additional insights}
\paragraph{Temporal attention pattern analysis}
\begin{figure*}[htbp]
    \centering
    \includegraphics[width=1\textwidth]{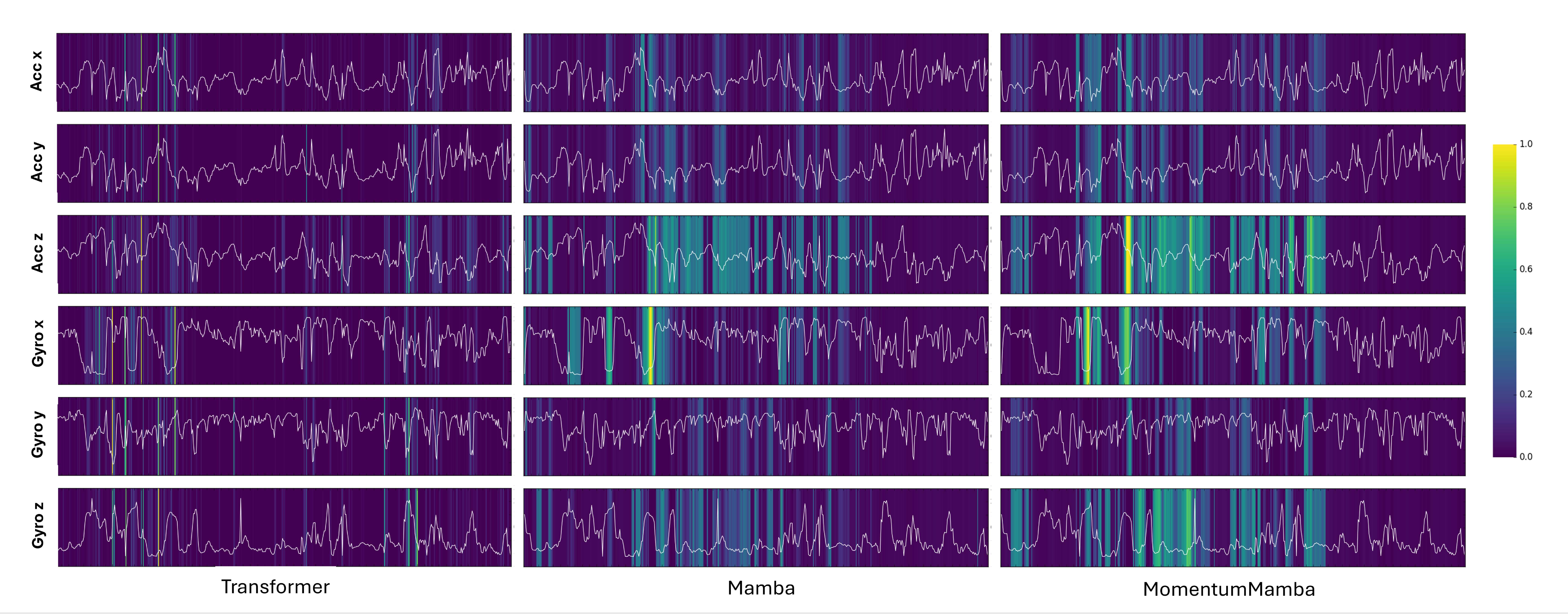}
    \caption{Two-Step Rescaling Explanations \cite{tsinterpret}: Comparative Saliency Analysis of Transformer, Mamba, and MomentumMamba Models on UESTC Dataset, where warmer colors (yellow) indicate higher feature importance and cooler colors (purple) indicate lower importance.}
    \label{fig:attention_heatmap}
\end{figure*}
To gain deeper insight into the internal mechanisms of different sequence modeling architectures, we visualize and compare the temporal saliency maps produced by three models: Transformer, Vanilla Mamba, and our proposed Momentum Mamba. Figure~\ref{fig:attention_heatmap}, each model processes the same six-dimensional inertial input, with the raw signal overlaid in white and the background heatmap indicating time-varying saliency.

The Transformer model tends to focus its attention on isolated, high-salience peaks typically centered around abrupt transitions or local extrema in the signal. These activations are sparse and vary significantly across input channels, indicating that the model attends selectively but inconsistently across features. Furthermore, early portions of the sequence often receive negligible attention, suggesting limited capacity for long-term memory retention.

Vanilla Mamba, in contrast, demonstrates smoother and more distributed attention compared to the Transformer. It allocates saliency across longer temporal spans, with more stable patterns in the mid- and late-sequence regions. However, its focus is still less consistent than that of Momentum Mamba, particularly in terms of early sequence sensitivity and multi-channel alignment. While Mamba benefits from input-conditioned state transitions and structured recurrence, it lacks an explicit mechanism to preserve information from past states beyond first-order dynamics.

Momentum Mamba shows a clear improvement in both breadth and continuity of temporal attention. The saliency maps reveal sustained activation across wide time windows, including the initial segments of the sequence. This indicates that Momentum Mamba is capable of retaining early-stage information and using it in downstream prediction--a critical feature for HAR tasks where initial motion cues can be subtle yet discriminative. Moreover, the saliency patterns are temporally smooth and spatially aligned across channels, implying a coordinated and stable latent representation.

We attribute this improvement to the incorporation of second-order momentum into the state update mechanism. By allowing the hidden state to evolve not only based on the current input but also on the velocity of past updates, Momentum Mamba effectively propagates information over time with greater persistence. This results in richer temporal representations and a more robust ability to model long-range dependencies in multichannel sensor data.

Overall, these visualizations provide qualitative evidence that Momentum Mamba captures temporal structure more effectively than both Transformer and Vanilla Mamba, supporting its superior performance in downstream HAR tasks.

\paragraph{Gradient flow analysis}
\begin{figure*}[htbp]
\centering
\includegraphics[width=0.8\textwidth]{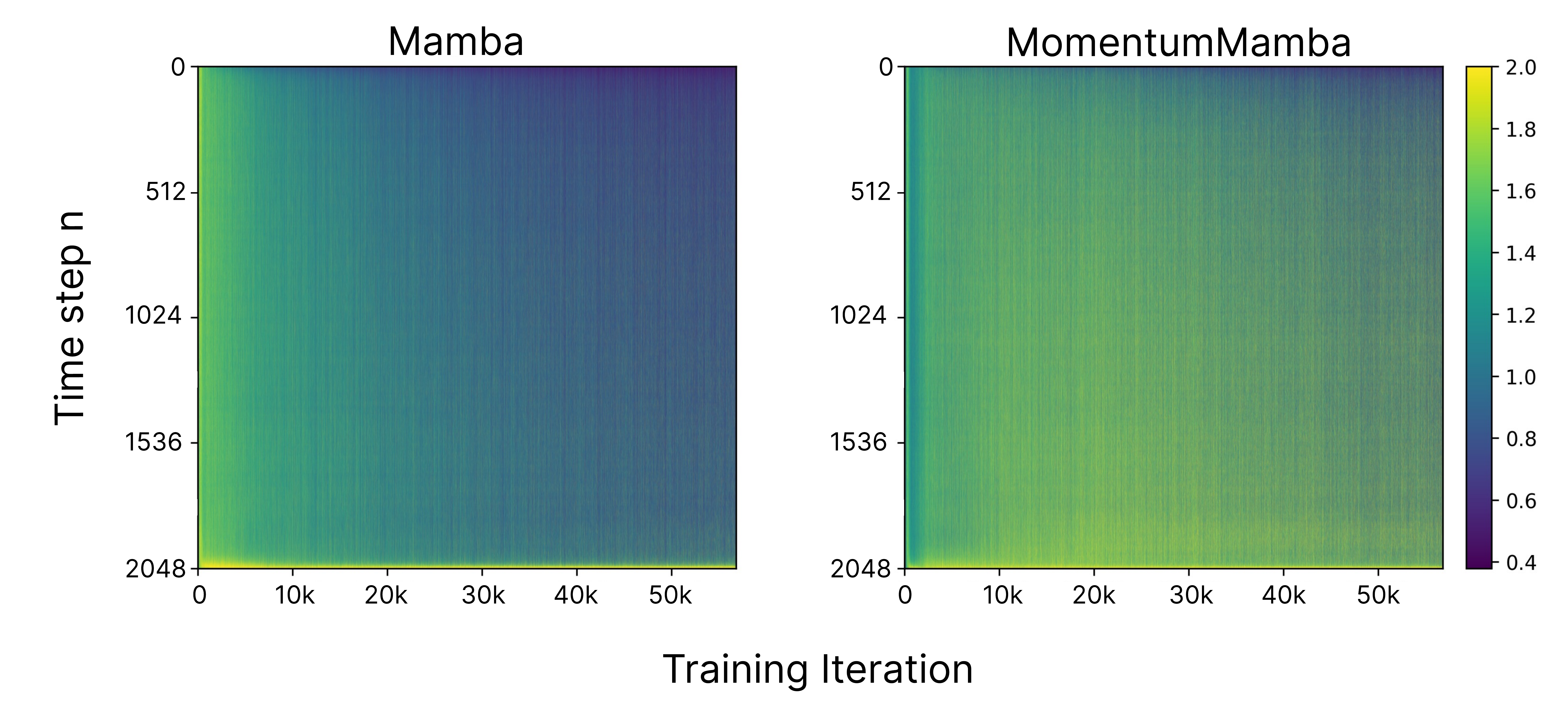}
\caption{$\ell_{2}$ norm of the gradients of the loss $\mathcal{L}$ w.r.t.\ the state vector $h_{t}$ at each time step $t$ for Vanilla Mamba (left) and MomentumMamba (right). MomentumMamba does not suffer from vanishing gradients.}
\label{fig:vanishing}
\end{figure*}

To better understand the optimization behavior of different recurrent state-space architectures, we examine the $\ell_{2}$ norm of the gradients of the loss $\mathcal{L}$ with respect to the hidden state $h_{t}$ across both time steps and training iterations (Fig.~\ref{fig:vanishing}). For Vanilla Mamba (left), gradient norms diminish rapidly as the time horizon increases. This pattern is symptomatic of the vanishing gradient problem, where information from early steps in long sequences fails to propagate back effectively during training. As a result, the model has limited ability to assign credit to distant dependencies, which ultimately constrains its capacity to capture extended temporal structure.

In contrast, Momentum Mamba (right) exhibits substantially more stable gradient magnitudes across both shallow and deep time steps, with relatively uniform patterns persisting throughout the training process. This observation indicates that the introduction of momentum into the state update mechanism directly improves gradient flow by creating an auxiliary pathway for information propagation. The second-order recurrence not only mitigates exponential decay in gradient norms but also allows informative updates from earlier time steps to persist over long horizons.

This behavior aligns closely with findings in MomentumRNN~\cite{nguyen2020momentumrnn}, where momentum-based recurrence was shown to counteract gradient shrinkage and maintain long-term credit assignment. By analogy, Momentum Mamba inherits these benefits within the structured state-space modeling framework, thereby providing a principled solution to one of the long-standing challenges in training deep sequential models. Crucially, the preservation of non-vanishing gradients ensures that the model can stably exploit both short-term variations and long-term trends in inertial sequences, directly contributing to the empirical performance improvements reported across MuWiGes, UESTC-MMEA-CL, and MMAct.

\paragraph{Hyperparameter sensitivity analysis}
\begin{figure*}[htbp]
\centering
\includegraphics[width=1\textwidth]{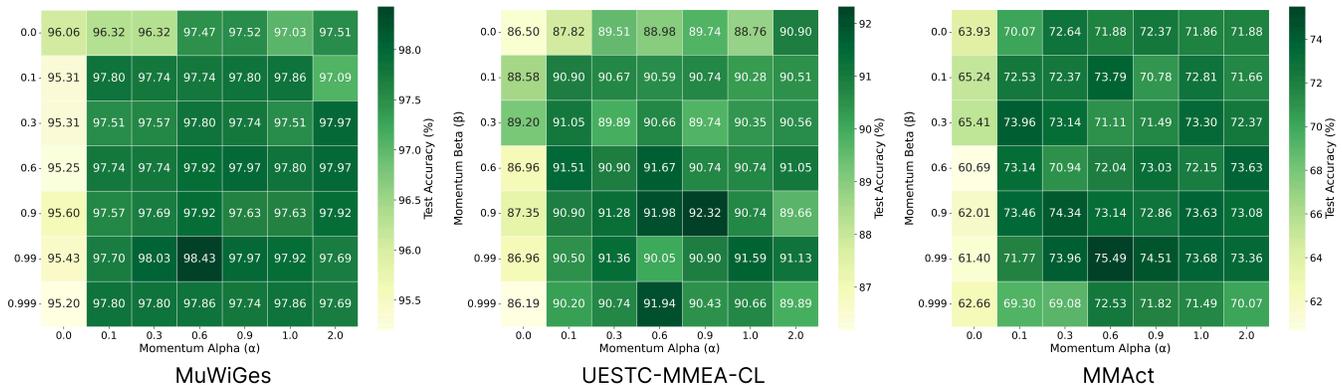}
\caption{Hyperparameter Grid Search Results for Momentum Mamba Model. Heatmap showing test accuracy (\%) across different combinations of momentum beta ($\beta$) and momentum alpha ($\alpha$) parameters. Each cell displays the test accuracy percentage achieved with the corresponding hyperparameter combination. The color scale ranges from lower accuracy (red) to higher accuracy (green), from the MuWiGes \cite{nguyen2023hand}, UESTC-MMEA-CL \cite{xu2023towards}, and MMAct \cite{kong2019mmact} datasets.}
\label{fig:beta_alpha_accuracy}
\end{figure*}

To evaluate the robustness of Momentum Mamba in different momentum settings, we performed a comprehensive grid search over two key hyperparameters: the momentum coefficient $\beta \in \{0.0, 0.1, 0.3, 0.6, 0.9, 0.99, 0.999\}$ and the input scaling factor $\alpha \in \{0.0, 0.1, 0.3, 0.6, 0.9, 1.0, 2.0\}$. Figure~\ref{fig:beta_alpha_accuracy} shows the test accuracy achieved on three benchmark HAR datasets: MuWiGes, UESTC-MMEA-CL and MMAct.

In the MuWiGes dataset, Momentum Mamba consistently achieved high accuracy across a wide range of hyperparameter settings. The highest accuracy of 98.43\% was attained at $(\beta=0.99, \alpha=0.6)$, with many other combinations achieving above 97.5\%. This indicates the model's robustness to hyperparameter variations and its ability to maintain strong performance even with minimal tuning.

For UESTC-MMEA-CL, which involves more subtle and fine-grained activities, the model achieved its peak performance of 92.32\% at $(\beta=0.9, \alpha=0.9)$. While performance degraded when $\beta$ was too small (e.g., $\beta=0.0$), moderate momentum coefficients (e.g., $\beta \in [0.6, 0.999]$) provided stable results above 90\%, indicating the importance of second-order memory in modeling fine-grained temporal variations.

On the more challenging MMAct dataset, the accuracy trend revealed a stronger dependence on tuning. Performance increased steadily with $\alpha$, particularly when paired with mid-to-high momentum ($\beta \in [0.6, 0.99]$). The highest test accuracy of 75.49\% was achieved at $(\beta=0.99, \alpha=0.6)$. In particular, when $\beta=0.0$, the accuracy plateaued below 71\%, reinforcing the benefit of incorporating momentum into the state update dynamics.

In general, these results demonstrate the followings.
\begin{itemize}
    \item Momentum Mamba benefits significantly from momentum-enhanced recurrence, particularly with $\beta$ in the range $[0.6, 0.999]$;
    \item The input scaling factor $\alpha$ complements $\beta$ by controlling the contribution of new input information to the momentum state;
    \item Properly chosen hyperparameters yield consistent improvements across datasets of varying complexity and motion dynamics.
\end{itemize}

These findings validate the utility of second-order dynamics in stabilizing the gradient flow and improving the model's ability to capture long-term dependencies in HAR tasks.

\section{Conclusion}
\label{sec:conclusion}
HAR with inertial sensors remains challenging due to noisy signals, limited spatial context, and strict resource constraints. To address these issues, we proposed Momentum Mamba, a momentum-augmented selective state-space model that enriches hidden-state evolution with a second-order dynamics pathway. This design mitigates vanishing gradients, enhances robustness against noise, and improves long-range temporal modeling.

We further introduced Complex Momentum Mamba for frequency-sensitive oscillatory memory and Adam Momentum Mamba for variance-aware adaptivity, extending the expressive capacity of state-space models. Comprehensive experiments on multiple HAR benchmarks confirmed that our models consistently outperform Transformer-based, RNN-based, and oscillatory SSM baselines, while preserving linear-time scalability and offering a favorable balance between accuracy and efficiency for real-time wearable deployment.

Looking forward, future work will focus on lightweight variants, adaptive trade-offs between accuracy and efficiency, and multimodal integration. Beyond HAR, momentum-augmented SSMs hold promise for a broad range of sequential domains such as biosignal analysis, robotics, speech, and multimodal temporal understanding, paving the way for scalable and interpretable sequence learning.

\bibliographystyle{unsrt}
\bibliography{references}

\end{document}